\documentclass[10pt]{iopart}
\usepackage{graphicx}

\begin{document}

%

\jl{6}

\title[Hollow sphere\ldots]{Hollow sphere, a flexible multimode
	Gravitational Wave antenna}

\author{J Alberto Lobo}

\address{Departament de F\'\i sica Fonamental, Universitat de
	Barcelona, Diagonal 647, 08028 Barcelona, Spain.}

\begin{abstract}
Hollow spheres have the same theoretical capabilities as the usual
solid ones, since they share identical symmetries. The hollow sphere
is however more flexible, as thickness is an additional parameter one
can vary to approach given specifications. I will briefly discuss
the more relevant properties of the hollow sphere as a GW detector
(frequencies, cross sections), and suggest some scenarios where it
can generate significant astrophysical information
\end{abstract}




\section{Introduction}

A Gravitational Wave (GW) is a time--dependent perturbation of the
background space-time geometry which propagates away from the source,
and which is generated by changes in the mass--energy distribution
of the latter. Due to the weakness of gravitation, the mathematical
description of GWs, so far as detection by a remote observatory is
concerned, can be very accurately accomplished by means of a
{\it linear approximation\/} in the field equations. More precisely,
if $\eta_{\mu\nu}$ is the (Lorentzian) background metric, the actual
metric $g_{\mu\nu}({\bf x},t)$ can be expediently split up as

\begin{equation}
 g_{\mu\nu}({\bf x},t) = \eta_{\mu\nu} + h_{\mu\nu}({\bf x},t)\ ,
 \qquad |h_{\mu\nu}|\ll 1
 \label{eq.1}
\end{equation}
where {\bf x} and $t\/$ are quasi--Lorentzian coordinates, and the
$h_{\mu\nu}({\bf x},t)$ are the GW perturbations.

Since GWs will of course be observed far from their source they can
be safely assumed to be {\it plane\/} across the detector's span.
In this case only the six so called {\it electric\/} components,
$R_{0i0j}({\bf x},t)$, of the Riemann tensor associated with~\eref{eq.1}
are not identically zero. Although General Relativity theory predicts
that at least four of these are zero, as explained in textbooks
\cite{mtw}, alternative theories of gravity (e.g.\ Brans--Dicke) are
generally less restrictive ---see~\cite{el73,will} for a complete
discussion of all the possibilities. It thus appears that detection
of GWs is a strong experimental test on which is the theory which
best describes gravity.

Currently operating Weber bars, as well as interferometric detectors,
such as {\sl LIGO\/} and {\sl VIRGO}, are however {\it single channel\/}
devices, i.e., they can only sense {\it one\/} linear combination of
the 5 quadrupole GW amplitudes ---the so called {\it antenna
pattern\/}~\cite{duti,mont}. A {\it network\/} of such detectors in
joint operation is thus necessary for a complete deconvolution of the
GW~signal, but this poses non-trivial practical problems of analysis,
because there are no two identical devices, noise characteristics
change from one to the other, etc.

A simple way out of such limitation is to use a {\it spherical detector},
because this is a natural {\it multimode}, or tensor, GW antenna
\cite{fw71,wp77,lobo}. The reason for this is that, having spherical
symmetry, an elastic solid has oscillation eigenmodes which come in
{\it degenerate} $l\/$-pole harmonics, the monopole ($l=0$) and
quadrupole ones efficiently coupling to the incoming GW~\cite{clo}.
By use of suitable sensor systems~\cite{jm93,mnras} it is theoretically
possible to completely {\it deconvolve\/} the GW signal, i.e., to
measure {\it all\/} six amplitudes which characterise a general
metric GW~\cite{will}.

A {\it hollow\/} sphere is an interesting variation of the solid
one~\cite{vega} which has a number of interesting capabilities as
a GW detector, and flexibility to implement them by adjusting the
thickness parameter. In this paper I intend to summarise the most
relevant features of an {\it ideal\/} hollow spherical GW antenna
(i.e., I will {\it assume\/} that it is technically possible to
build a working system), and see what interesting physics could
be done with it.

\section{Response and sensitivity}

\begin{figure}
\centering
\includegraphics[width=10cm]{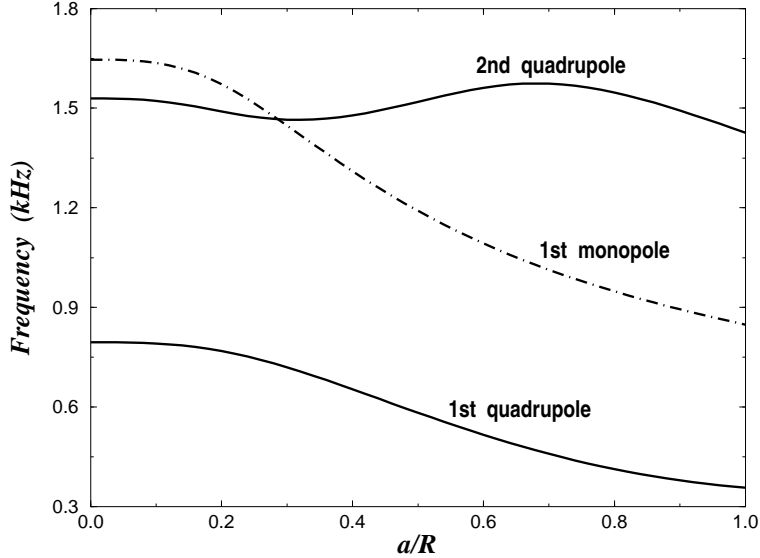}
\caption{The first two quadrupole (solid lines) and first monopole 
(dash-dotted) frequencies of a hollow sphere of a CuAl alloy, 3 metres
of outer diameter, as a function of the ratio between the inner radius,
$a$, and the outer radius, $R$.}
\label{fig.1}
\end{figure}

As usual (see \cite{lobo} for a thorough discussion), the solid's
response to a GW excitation is given by the solution, with appropriate
boundary conditions, to the partial differential equation

\begin{equation}
\varrho\;\frac{\partial^2 {\bf u}}{\partial t^2} - \mu\,\nabla^2 {\bf u}
  - (\lambda+\mu)\,\nabla(\nabla{\bf\cdot}{\bf u}) = {\bf f}({\bf x},t)
\label{eq.2}
\end{equation}

where $\varrho$, $\lambda$ and $\mu\/$ are the density and elastic
Lam\'e coefficients, respectively, and

\begin{equation}
 {\bf f}({\bf x},t) = {\bf f}^{(00)}({\bf x})\,g^{(00)}(t)\ +\ 
  \sum_{m=-2}^2\,{\bf f}^{(2m)}({\bf x})\,g^{(2m)}(t)
 \label{eq.3}
\end{equation}

is the {\it tidal\/} GW driving density of force. The hollow sphere's
response is then given by

\begin{equation}
 \hspace*{-1.4 cm}
 {\bf u}({\bf x},t) = \sum_{n=1}^\infty\,\frac{a_{n0}}{\omega_{n0}}
   \,{\bf u}_{n00}({\bf x})\,g_{n0}^{(00)}(t)\ +\ 
   \sum_{n=1}^\infty\,\frac{a_{n2}}{\omega_{n2}}\,\left[\sum_{m=-2}^2
   \,{\bf u}_{n2m}({\bf x})\,g_{n2}^{(2m)}(t)\right]
 \label{eq.4}
\end{equation}

where $\omega_{nl}\/$ are the oscillation eigenfrequencies of the
${\bf u}_{nlm}({\bf x})$ eigenmode, $a_{nl}\/$ are overlap coefficients
and $g_{nl}^{(lm)}(t)$ is the convolution between the signal
$g^{(lm)}(t)$ and the detector mode. Eigenmode wave functions
${\bf u}_{nlm}\/$ are somewhat complicated functions which are
calculated in detail in reference~\cite{vega}, while the frequencies
of the first two quadrupole and first monopole harmonics are plotted
in figure~\ref{fig.1} for a conceivable materials and dimensions of
the system.

\begin{figure}
\centering
\includegraphics[width=10cm]{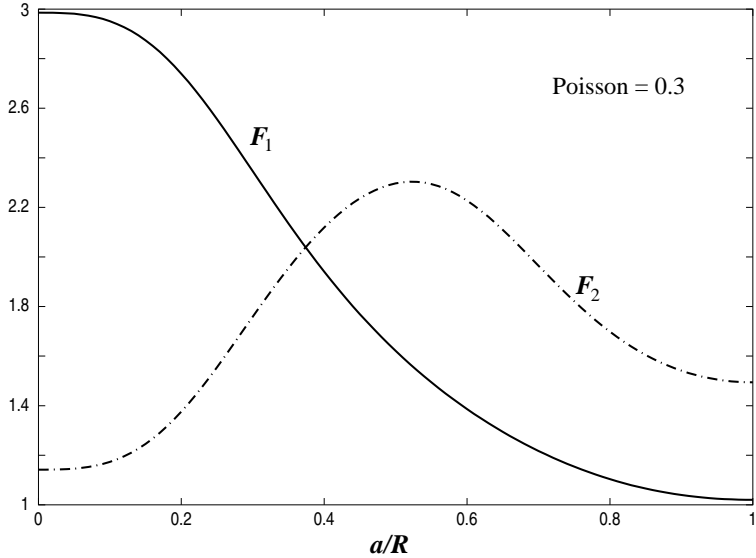}
\caption{The first two quadrupole cross section coefficients of
equation~\protect\eref{eq.5} for hollow spheres of a CuAl alloy.
In abscissae the ratio between the inner radius, $a$, and the
outer radius, $R$. This plot is of course independent of the
antenna's absolute size.}
\label{fig.2}
\end{figure}

The most relevant parameter characterising the solid's sensitivity
to GWs is the {\it cross section\/} for the absorption of incoming
GW energy. As shown in reference~\cite{vega}, the following formula
holds for quadrupole modes:

\begin{equation}
 \sigma_n \equiv
 \sigma_{\rm abs}(\omega_{n2})= F_n(\sigma_P;a/R)\,\frac{GMv_s^2}{c^3}
 \label{eq.5}
\end{equation}

where $v_s$ is the speed of sound in the material, $\sigma_P\/$ its
Poisson ratio, and $F_n(\sigma_P;a/R)$ is a dimensionless parameter
associated with the $n\/$-th quadrupole mode of the hollow sphere.
A plot of the first two $F_n\/$'s is given in figure~\ref{fig.2}.
The crossing between the curves happens at $a/R=0.375$, which means
a hollow sphere is a more efficient GW energy absorber in its second
mode than in the first, which favours in turn higher frequency signal
reception.

\section{Physics with a hollow sphere}

Let us now briefly assess which sources of GWs can be visible with a
hollow spherical detector, under the ideal assumption that technological
problems are solved to the effect that the detector's quantum limit
has been reached. If such is the case then peak sensitivity is given
by~\cite{vega}

\begin{equation}
 S_h(\omega_{n}) = \frac{G\hbar}{c^3}\;\frac{4\pi\beta_n}{\sigma_n}
 \label{eq.6}
\end{equation}

where $\omega_{n}\/$ is a quadrupole resonance frequency, and $\beta_n\/$
is the efficiency of the coupling between mechanics and electronics in that
mode; the detector's bandwidth is $\Delta\omega_n=\beta_n\omega_n$.

Take for example a hollow sphere of CuAl, 3 metres of outer diameter
and 22 centimetres thick ($a/R=0.15$), weighing 40 tons; the first
two quadrupole frequencies are then 400 and 1500 Hz, respectively,
and the peak sensitivity is essentially identical in both modes,
reaching $10^{-46}$ Hz$^{-1}$ for a $\beta\/$ factor of 0.1. A
system like this would be able to see a millisecond burst with an
amplitude of $10^{-22}$ metres/metre, and a monochromatic signal as
weak as $10^{-27}$ metres/metre at the first resonant frequency if
integrated for one year. Such sensitivities would most likely enable
the observations of supernova explosions in galaxies out to the Virgo
cluster (at a rate of a few per year), and a significant sample of
the galactic pulsar population.

A larger detector, 6 metres in outer diameter, 25 centimetres thick,
200 tons of CuAl alloy, has a minimum spectral density of noise of
$3\times 10^{-47}$ Hz$^{-1}$ at its first resonance (190 Hz), and
$2\times 10^{-47}$ Hz$^{-1}$ at the second (750 Hz). Such a system
can be used for {\it chirp\/} signal detection by a robust method
first described for spherical detectors by Coccia and Fafone~\cite{cf}.
The idea is conceptually simple: the GW signal generated e.g.\ by
a compact binary system nearing coalescence sweeps an ascending
frequency range which eventually overlaps with the sensitivity
bands of the detector. If the binary remains undisrupted~\cite{clark}
while in the frequency intervals between the first two just quoted
resonances then the antenna's response will peak at those frequencies,
with a signal to noise ratio given by

\begin{equation}
 SNR = \frac{5\pi 2^{1/3}}{24}\;\frac{(GM_c)^{5/3}}{c^3}\;
 \frac{\Delta\omega_n}{S_h(\omega_n)}\;\frac{1}{r^2\,\omega_n^{7/3}}
 \label{eq.7}
\end{equation}

where $r\/$ is the distance to the source, $\omega_n\/$ is a detector
quadrupole resonance frequency, and $M_c\/$ is the {\it mass parameter}

\begin{equation}
 M_c\equiv(m_1 m_2)^{3/5}(m_1 + m_2)^{-1/5}
 \label{eq.8}
\end{equation}

of the chirp signal. Let now $\tau_1$ and $\tau_2$ be the times at
which the the signal reaches the detector with frequencies $\omega_1$
and $\omega_2$, respectively. The mass parameter can then be estimated
by the formula

\begin{equation}
 M_c = \frac{5^{3/5}}{2^{16/5}}\;\frac{c^3}{G}\,\left(
 \frac{\omega_2^{-8/3}-\omega_1^{-8/3}}{\tau_2-\tau_1}\right)^{\!3/5}
 \label{eq.9}
\end{equation}

in which the lowest post-Newtonian approximation for the waveform has
been assumed~\cite{cf}. The strong point of this {\it double passage\/}
method is its {\it robustness\/} relative to the details of the signal's
{\it phase\/}~\cite{cutler,tho}, which considerably complicates the
performance of the {\it matched filter\/} in extracting the signal
from the noise in an interferometric detector~\cite{dhur}. It thus
appears that both detectors, a sphere and an interferometer, can
constructively cooperate in the identification of compact binary
coalescences, as the sphere can provide a good estimate of the mass
parameter which will simplify and enhance the efficiency of the bank
of filters needed in an interferometer~\cite{dhur}.

The hollow sphere is particularly suited for this, as it can be made to
have low resonance frequencies, and this favours SNR because the signal
stays longer in the lower frequencies ---gravitational bremsstrahlung
tends to accelerate the inspiral as the binary's period of revolution
diminishes. This fact is clearly reflected by equation~\eref{eq.7},
where SNR is seen to strongly depend on the 7/3 inverse power of
frequency.

{\it Cosmic backgrounds\/} of GWs can also be observed with hollow
spheres. Such signals, due to their {\it stochastic\/} nature, are
impossible to tell from the detector noise. A minimum of two detectors,
whose outputs are cross-correlated, must be used in order to see a
common signal which can be identified by a distinctive spectral
density, $S_{\rm GW}(\omega)$. With unit SNR, a pair of identical
hollow spheres could set a good limit on the density parameter of
GWs in the Universe~\cite{als,vega}:

\begin{equation}
 \hspace*{-2.3 cm}
 \Omega_{\rm GW}\simeq 10^{-9}\,\times\,
 \left(\frac{f_{n}}{200\ {\rm Hz}}\right)^{\!3}\,
 \left(\frac{S^{(1)}_h(f_n)\,S^{(2)}_h(f_n)}
  {10^{-96}\ {\rm Hz}^{-2}}\right)^{\!1/2}\,
 \left(\frac{20\ {\rm Hz}}{\Delta f_n}\right)^{\!1/2}\,
 \left(\frac{1\ {\rm year}}{\tau}\right)^{\!1/2}
 \label{eq.10}
\end{equation}

where $\tau\/$ is the integration time and $\Delta f_n$ the detector's
bandwidth around $f_n$.

It has recently been conjectured by Amelino-Camelia~\cite{amelino},
on the basis of dimensional arguments, that a cosmic background of
GWs could have, in first approximation, a flat spectrum with spectral
density

\begin{equation}
 S_{\rm GW}(\omega)\sim\frac{\ell_{\rm Planck}}{c} = \left(
 \frac{G\hbar}{c^5}\right)^{1/2}\sim 5\times 10^{-44}\ {\rm Hz}^{-1}
\end{equation}

It is unclear whether this is a realistic assumption~\cite{pia}; if it
were however it would mean that Planck scale scale physics would become
accessible to the next generation of GW detectors, and hollow spheres
would enable the observation of such phenomena with signal to noise
ratios of the order of about 100.

\section{Closing remarks}

As already mentioned, spherical elastic bodies are natural {\it multimode\/}
GW antennas. The reason for this is the degeneracy of their quadrupole
oscillation eigenmodes, which ideally match the structure of GWs. Both
a solid and a hollow sphere share this property, of course, and in this
paper I have briefly considered some of the more interesting properties
of the latter, related to its sensitivity and subsequent capabilities as
a powerful GW telescope. I have deliberately left aside the issues of
how it could be cast and suspended in practice, as well as questions
of how a quantum limited transducer system can be coupled efficiently
to the body of the antenna. The paper is thus intended to give a
consistent flavour of the best performance one can possibly aim at
with such a detector.

There being the possibility to adjust the spherical shell's thickness,
the hollow sphere is a more flexible device than a solid one to meet
given specifications. In fact, recent research work~\cite{dual} shows
that it is advantageous to use a hollow and a solid sphere placed in
a joint concentric layout to produce a GW detector which has enhanced
sensitivity over a frequency band between the resonances of the inner
solid body and the outer hollow one. By an optimised choice of the
latter's thickness, and by means of optical Fabry-Perot cavities
acting as non-resonant transducers, such a {\it nested\/} sphere
system appears to be a promising device for GW signal detection in
the kHz region in bandwidths of several hundred Hz, where it beats
the prospects of the very sensitive earth based interferometric GW
detectors. And, let me stress this once more, all of it with the
ability to potentially deconvolve the five quadrupole amplitudes
with a single system.

\section*{Acknowledgements}

I acknowledge with thanks support from the Spanish Ministry of
Education, contract code {\tt BFM2000-0604}.

\end{document}